\documentclass[floatfix,twocolumn]{aastex631}
\usepackage{threeparttable, longtable, threeparttablex, booktabs, url, amsmath}
\usepackage{verbatim}
\usepackage{comment}
\usepackage{float}
\usepackage{multirow}

\graphicspath{{./}{figures/}}
\RequirePackage[normalem]{ulem} 
\RequirePackage{color}\definecolor{RED}{rgb}{1,0,0}\definecolor{BLUE}{rgb}{0,0,1} 

\newcommand{\be}{\begin{eqnarray} }
\newcommand{\ee}{ \end{eqnarray} }

\shorttitle{}
\shortauthors{Feng et al.}
\graphicspath{{./}{figures/}}
\begin{document}

\title{The Real and Pseudo Dispersion Measures of FRB~20220912A}

\correspondingauthor{Yi Feng, Y.-K. Zhang, D. Li}
\email{yifeng@zhejianglab.org, ykzhang@nao.cas.cn, dili@tsinghua.edu.cn}

\author[0000-0002-0475-7479]{Yi Feng}
\affil{Research Center for Computational Earth and Space Science, Zhejiang Laboratory, Hangzhou 311100, China}
\affil{Institute for Astronomy, School of Physics, Zhejiang University, Hangzhou 310027, China}

\author[0000-0002-7420-9988]{Dengke Zhou}
\affil{Research Center for Computational Earth and Space Science, Zhejiang Laboratory, Hangzhou 311100, China}

\author[0000-0002-8744-3546]{Yongkun Zhang}
\affil{National Astronomical Observatories, Chinese Academy of Sciences, Beijing 100101, China}

\author[0000-0003-3010-7661]{Di Li}
\affiliation{New Cornerstone Science Laboratory, Department of Astronomy, Tsinghua University,Beijing 100084, China}
\affiliation{National Astronomical Observatories, Chinese Academy of Sciences, Beijing 100101, China}

\author[0000-0001-9956-6298]{Jianhua Fang}
\affil{Research Center for Computational Earth and Space Science, Zhejiang Laboratory, Hangzhou 311100, China}

\author[0000-0002-9579-6739]{Jiaying Xu}
\affil{Research Center for Computational Earth and Space Science, Zhejiang Laboratory, Hangzhou 311100, China}

\author[0000-0002-4968-2466]{Chenyuan Xu}
\affil{Research Center for Computational Earth and Space Science, Zhejiang Laboratory, Hangzhou 311100, China}

\author[0000-0001-5649-2591]{Jintao Xie}
\affil{School of Computer Science and Engineering, Sichuan University of Science and Engineering, Yibin 644000, China}



\begin{abstract}
Fast radio bursts (FRBs) are millisecond-duration radio transients. As they propagate through the interstellar medium, they interact with free electrons, resulting in dispersion. The corresponding dispersion measure (DM) is referred to as the real DM (DM$_{\rm real}$). In practice, however, the dispersion measure derived from modeling (DM$_{\rm model}$) is often contaminated by intrinsic burst morphology, giving rise to a pseudo DM component (DM$_{\rm pseudo} = {\rm DM}_{\rm model} - {\rm DM}_{\rm real}$). In this work, we focus on the highly active repeating FRB~20220912A and utilize its microshots---extremely short-duration (typically tens of microseconds), broadband emissions---to investigate its DM$_{\rm real}$ and DM$_{\rm pseudo}$. We adopt two assumptions: first, that FRB~20220912A resides in a non-magneto-ionic environment and that its DM$_{\rm real}$ variation is smaller than $10^{-2}$\,pc\,cm$^{-3}$ over a few years; and second, that microshots have a negligible intrinsic morphological time delay. By identifying two new microshots and combining them with previously reported ones, we find that all four microshots exhibit remarkably consistent DM values over a one-month timescale, with an average of $219.380 \pm 0.004\,\mathrm{pc\,cm^{-3}}$. We define this value as the DM$_{\rm real}$ of FRB~20220912A. We further show that bright, narrow bursts with a width of less than 2\,ms also yield DM estimates consistent with the microshot-based DM$_{\rm real}$. A survey of five repeating FRBs reveals that DM$_{\rm pseudo}$ is a common phenomenon, with variations typically spanning a range of approximately $10\,\mathrm{pc\,cm^{-3}}$ at 1.2\,GHz. These findings highlight the importance of accounting for morphological contributions in DM interpretation and demonstrate that microshots and narrow bursts are powerful tools for probing DM$_{\rm real}$.
\end{abstract}

\section{Introduction}
\label{sec:intro}
Fast radio bursts (FRBs) are bright, millisecond-duration radio transients, first discovered by \cite{2007Sci...318..777L}. Despite extensive observational efforts, their physical origins and emission mechanisms remain uncertain \citep{zhangreview2023}. Among the roughly 4,000 known FRBs, about 100 sources exhibit repeated bursts and are classified as repeating FRBs\footnote{See \url{https://blinkverse.zero2x.org}} \citep{blinkverse}. This repeating behavior enables detailed, multi-epoch investigations into burst properties---such as spectral structure, temporal evolution, and polarization---offering critical clues to understand their progenitor environments and radiative processes.

The dispersion measure (DM) is defined as:
\begin{equation}
\text{DM} = \int_{0}^{d} n_e(l) \, dl,
\label{eq:dm_definition}
\end{equation}
where \( d \) is the distance to the source in parsecs, \( l \) is the line-of-sight position, and \( n_e \) is the free electron density along the line of sight. We refer to the DM defined by Eq.~\ref{eq:dm_definition} as the real DM (DM\(_{\text{real}}\)).
When the FRB signal propagates through the interstellar medium (ISM), interactions with free electrons cause dispersion, characterized by a frequency-dependent time delay \(\Delta t_{\text{ISM}}\):
\begin{equation}
\Delta t_{\text{ISM}} = \mathcal{D} \times \frac{\text{DM}_{\text{real}}}{f^2},
\label{eq:plasma_delay}
\end{equation}
where $f$ is the observing frequency, and \(\mathcal{D} = (4.148808 \pm 0.000003) \times 10^3 \, \mathrm{MHz}^2 \, \mathrm{pc}^{-1} \, \mathrm{cm}^3 \, \mathrm{s}\) is the dispersion constant.
DM\(_{\text{real}}\) is a fundamental quantity associated with FRB sources. It is often used to determine whether an FRB is located outside the Milky Way and to estimate its distance. A large host DM\(_{\text{real}}\), or variations in DM\(_{\text{real}}\), may trace the magneto-ionic environment surrounding the FRB (\citealt{niu22}, \citealt{niu26}, \citealt{feng25}, \citealt{2026ApJ...996L..16M}).

\begin{figure*}[htp]
    \centering
    \includegraphics[scale=0.42]{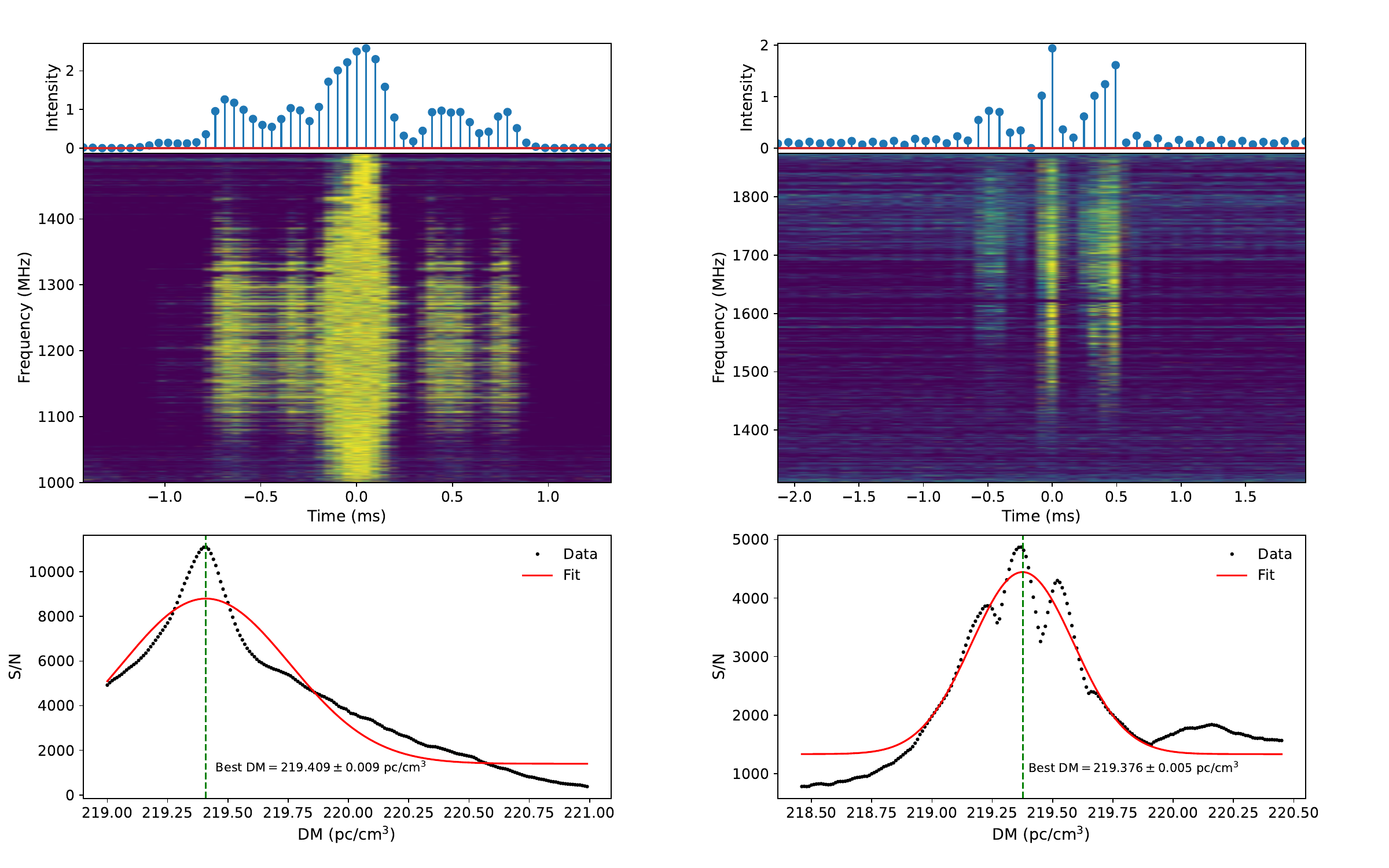}
    \caption{Two microshots are shown side by side. For each event, the top panel shows the frequency-averaged pulse profile, the middle panel displays the dynamic spectrum (frequency versus time), and the bottom panel presents the DM–S/N curve. The red line indicates the best-fit model consisting of a Gaussian plus a constant baseline, while the green dashed line marks the best-fitting DM value.}
    \label{fig:microshot}
\end{figure*}

Determining DM$_{\text{real}}$ is non-trivial, as the intrinsic morphology of FRBs is complex (\citealt{2019ApJ...876L..23H}, \citealt{2022RAA....22l4001Z}, \citealt{2025ApJ...979..160S}, \citealt{2026ApJ...998..276Z}). Some morphological features manifest as additional time delays, such that the total time delay is $\Delta t_{\text{total}} = \Delta t_{\text{ISM}} + \Delta t_{\text{mor}}$, where $\Delta t_{\text{mor}}$ is the delay due to burst morphology.
In practice, neither $\Delta t_{\text{ISM}}$ nor $\Delta t_{\text{mor}}$ is known a priori. The total time delay is therefore modeled to derive a dispersion measure, referred to as DM\(_{\text{model}}\), using tools such as \texttt{DM\_phase}\footnote{\url{https://github.com/danielemichilli/DM_phase} } \citep{2019ascl.soft10004S}. For FRBs with non-zero $\Delta t_{\text{mor}}$, DM$_{\text{model}}$ and DM$_{\text{real}}$ are generally not equal. We define the pseudo DM as the difference between the two: $\text{DM}_{\text{pseudo}} = \text{DM}_{\text{model}} - \text{DM}_{\text{real}}$. Theoretically, the intrinsic morphology, described by $\Delta t_{\text{mor}}$, is related to $\text{DM}_{\text{real}}$ via $\Delta t_{\text{mor}} = \Delta t_{\text{total}} - \Delta t_{\text{ISM}} = \Delta t_{\text{total}} - \mathcal{D} \times \frac{\text{DM}_{\text{real}}}{f^2}$. Thus, knowledge of the real DM is critical for measuring many other attributes related to intrinsic morphology, such as burst widths, peak flux densities, and scattering timescales. However, $\text{DM}_{\text{real}}$ is not known a priori. In practice, to reconstruct the intrinsic morphology, $\Delta t_{\text{mor}}$ is calculated using $\Delta t_{\text{total}} - \Delta t_{\text{ISM}}$ under the assumption that $\text{DM}_{\text{real}} = \text{DM}_{\text{model}}$.

FRB~20220912A is an extremely active repeating FRB, likely residing in a non-magneto-ionic environment (\citealt{2023ApJ...955..142Z}, \citealt{2024ApJ...974..296F}). It was first discovered by the Canadian Hydrogen Intensity Mapping Experiment (CHIME) \citep{2022ATel15679....1M} and subsequently followed up by telescopes worldwide, including the Deep Synoptic Array (DSA-110) \citep{2022ATel15716....1R}, the Big Scanning Antenna (BSA) \citep{2022ATel15713....1F}, the Robert C. Byrd Green Bank Telescope (GBT) \citep{2024ApJ...974..296F}, the Effelsberg 100-m Telescope \citep{2022ATel15727....1K}, the Five-hundred-meter Aperture Spherical Radio Telescope (FAST) \citep{2023ApJ...955..142Z}, the Allen Telescope Array \citep{2024MNRAS.52710425S}, the Nancay Radio Telescope (NRT) (\citealt{2023MNRAS.526.2039H}, \citealt{2024MNRAS.534.3331K}) and the upgraded Giant Metrewave Radio Telescope (uGMRT) \citep{2025arXiv251221889K}.
Its rotation measure (RM), measured to be near zero (\(\approx 0\) rad m\(^{-2}\)), remained relatively stable over a timescale of several months. Microshots in FRBs are defined as extremely short-timescale (typically tens of microseconds), broadband structures. Microshots of FRB~20220912A were first reported by \cite{2023MNRAS.526.2039H}.

To determine DM$_{\text{real}}$ of FRB~20220912A, we make two assumptions:
\begin{itemize}
\item[(A1)] FRB~20220912A is in a non-magneto-ionic environment, and its $\mathrm{DM_{real}}$ variation is smaller than $10^{-2}$ pc cm$^{-3}$ over a few years;
\item[(A2)] microshots have $\Delta t_{\text{mor}} = 0$, such that $\text{DM}_{\text{model}} = \text{DM}_{\text{real}}$.
\end{itemize}
These assumptions are motivated by the source's reported environmental stability (\citealt{2023ApJ...955..142Z}, \citealt{2024ApJ...974..296F}) and the expectation that microshots, being extremely narrow, are unlikely to possess significant intrinsic morphological delays.
Combining assumptions (A1) and (A2), microshots should exhibit a nearly constant DM$_{\text{model}}$.
In \cite{2023MNRAS.526.2039H}, two microshots separated by approximately three days were reported with remarkably consistent DM$_{\text{model}}$ values (219.356 $\pm$ 0.012 pc cm$^{-3}$ and 219.377 $\pm$ 0.009 pc cm$^{-3}$). This consistency makes it plausible that our assumptions are correct.

In this study, we identify two additional microshots from FRB~20220912A using data from FAST and GBT. Under assumptions (A1) and (A2), we use these microshots to determine DM\(_{\text{real}}\). The resulting DM\(_{\text{model}}\) values are mutually consistent over a timescale of approximately one month, providing strong evidence that our assumptions are valid.
We describe the identification of microshots and the determination of DM\(_{\text{real}}\) in Section~\ref{sec:dm}. In Section~\ref{sec:dis}, we test whether DM\(_{\text{model}} = \text{DM}_{\text{real}}\) holds for bright narrow bursts of FRB~20220912A and examine \(\text{DM}_{\text{pseudo}}\) in other repeating FRBs. We present our conclusions in Section~\ref{sec:con}.

\begin{figure*}[htbp]
    \centering
    \includegraphics[scale=0.6]{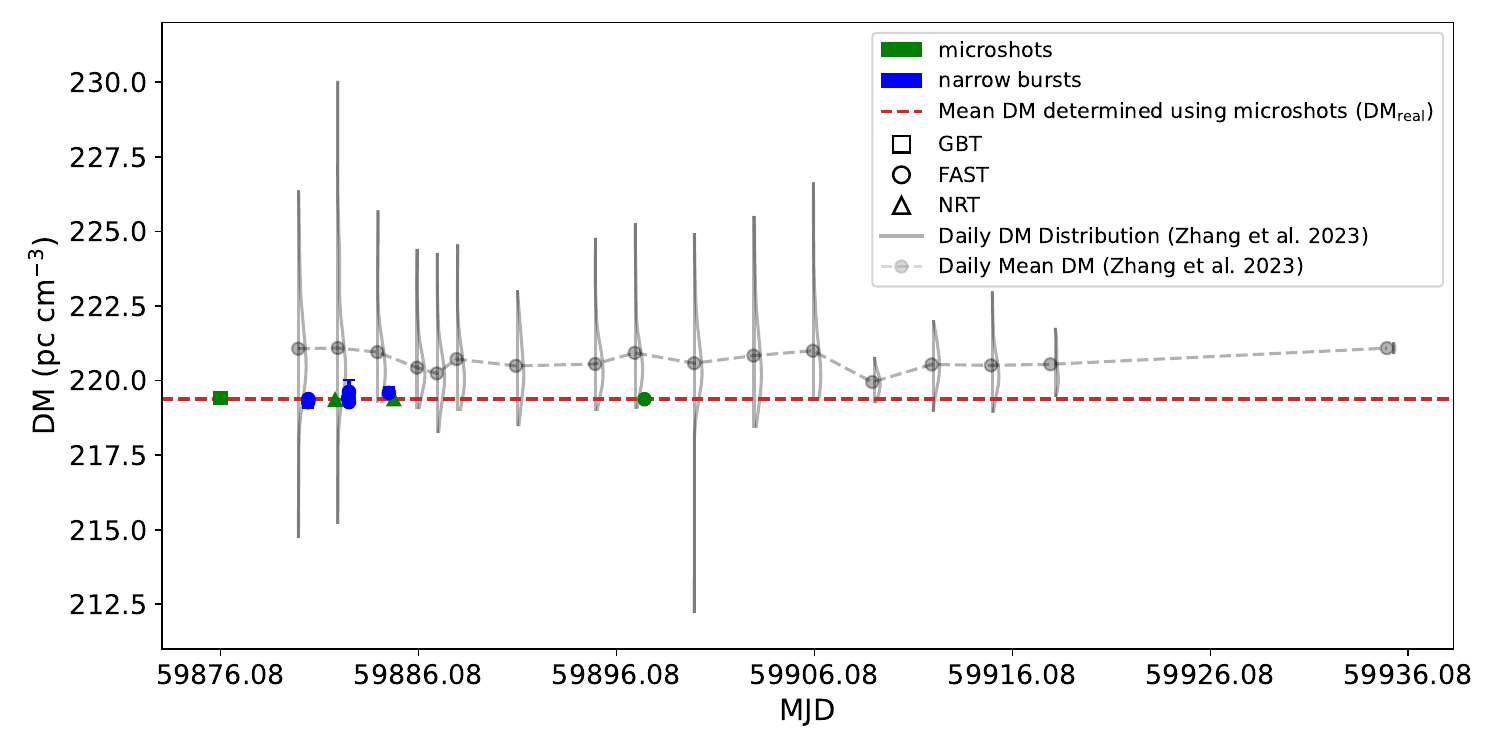} 
    \caption{DM$_{\text{model}}$ versus MJD for observations with GBT (squares), FAST (circles), and NRT (triangles). Green and blue markers indicate microshots and narrow bursts, respectively. Error bars denote 1$\sigma$ uncertainties. The dashed red horizontal line represents the weighted mean DM of the microshots, which we adopt as the DM$_{\text{real}}$. Right-hand black shaded distributions show the DM probability density functions for each day, estimated via Gaussian kernel density estimation from the measurements of \cite{2023ApJ...955..142Z}. Gray dashed lines with circular markers indicate the daily mean DM from the same reference. Only measurements from \cite{2023ApJ...955..142Z} with DM uncertainties $<$ 0.5\,pc\,cm$^{-3}$ are included, yielding 766 data points.}
    \label{fig:dm_vs_mjd}
\end{figure*}

\section{DM$_{\text{real}}$ Determination}
\label{sec:dm}
Using data from \cite{2023ApJ...955..142Z} and \cite{2024ApJ...974..296F}, we identify two additional microshots. In \cite{2024ApJ...974..296F}, 128 bursts were detected over 1.4 hours with the GBT. The data were coherently dedispersed at a DM of 219.46\,pc cm$^{-3}$ and recorded in the PSRFITS standard format \citep{2004PASA...21..302H}. Full-Stokes spectra were captured every 81.92\,$\mu$s with 0.195\,MHz-wide channels. Further details can be found in \cite{2024ApJ...974..296F}. Due to the sampling time of 81.92\,$\mu$s, microshots cannot be fully resolved, and typically, a microshot consists of one or two samples. We identify one microshot in the data from \cite{2024ApJ...974..296F}. In \cite{2023ApJ...955..142Z}, 1076 bursts were detected over 8.67 hours. The data were recorded in FITS format with a time resolution of 49.152\,$\mu$s, covering a frequency bandwidth from 1 to 1.5\,GHz with 4096 frequency channels. Details can be found in \cite{2023ApJ...955..142Z}. The data were not coherently dedispersed. The DM smearing at 1.2\,GHz is approximately 130\,$\mu$s. Consequently, the microshot cannot be fully resolved and consists of approximately three samples. We identify one microshot in the data of \cite{2023ApJ...955..142Z}.

We show the two newly discovered microshots in Figure~\ref{fig:microshot}. For each microshot, the top panel shows the frequency-averaged pulse profile, obtained by averaging the intensity across all frequency channels after baseline normalization. The middle panel displays the dynamic spectrum (frequency versus time), highlighting the fine temporal and spectral structure of the microshot. To determine the DM precisely, we followed the method proposed in \cite{2023MNRAS.526.2039H}. For a series of trial DMs around an initial guess, the data were fractionally dedispersed at sub-sample resolution, shifting each frequency channel according to the expected dispersive delay. The signal-to-noise ratio (S/N) was then computed within a narrow time window around the peak of each microshot, generating individual DM–S/N curves. The bottom panel shows the averaged DM–S/N curve over all considered microshots, which reduces statistical fluctuations and provides a more robust estimate of DM. A Gaussian plus constant baseline was fitted to this averaged curve to derive the best-fitting DM and its uncertainty. In the figure, the red line represents the Gaussian fit, and the green dashed line marks the resulting DM value. This procedure enables high-precision DM measurements for narrow microshots while minimizing biases from frequency-dependent noise and temporal smearing.

\begin{table*}[htbp]
\centering
\begin{threeparttable}
\caption{Parameters of the four microshots: (1) microshot number; (2) survey or reference; (3) detection time in Modified Julian Date (MJD); (4) DM$_{\text{model}}$ in units of pc\,cm$^{-3}$.}
\label{tab:microshots}
\begin{tabular}{cccc}
\hline
No. & Survey & MJD & DM$_{\text{model}}$ \\
    &        &     & (pc\,cm$^{-3}$) \\
\hline
1 & This work & 59876.07967444$^{a}$ & 219.409 $\pm$ 0.009 \\
2 & \cite{2023MNRAS.526.2039H} & 59881.86563025$^{b}$ & 219.356 $\pm$ 0.012 \\
3 & \cite{2023MNRAS.526.2039H} & 59884.83394050$^{b}$ & 219.377 $\pm$ 0.009 \\
4 & This work & 59897.49704621$^{a}$ & 219.376 $\pm$ 0.005 \\
\hline
\end{tabular}

\begin{tablenotes}
\item[a] Barycentric arrival time at 1.5 GHz.
\item[b] The arrival time is not barycenter corrected.
\end{tablenotes}

\end{threeparttable}
\end{table*}

The resulting DM\(_{\text{model}}\) values are 219.409 \(\pm\) 0.009\,pc\,cm\(^{-3}\) and 219.376 \(\pm\) 0.005\,pc\,cm\(^{-3}\), respectively. Table~\ref{tab:microshots} presents all DM measurements of microshots from FRB~20220912A, which includes the two newly identified microshots and two previously reported ones from \cite{2023MNRAS.526.2039H}. The average DM of these four microshots is \(219.380 \pm 0.004\,\mathrm{pc\,cm^{-3}}\). All four microshots are consistent with this average value within their uncertainties. According to assumptions (A1) and (A2), we adopt this average value as the real DM, i.e., \(\mathrm{DM}_{\text{real}} = 219.380 \pm 0.004\,\mathrm{pc\,cm^{-3}}\).
Figure~\ref{fig:dm_vs_mjd} displays the DM\(_{\text{model}}\) values of these four microshots along with the inferred DM\(_{\text{real}}\). The time interval between the two newly discovered microshots is nearly one month, an order of magnitude larger than that between the two microshots reported in \cite{2023MNRAS.526.2039H}, further supporting our two assumptions.

\section{DISCUSSIONS}\label{sec:dis}
\subsection{narrow bursts} \label{subsec:narrow}
While the use of microshots to estimate $\mathrm{DM}_{\text{real}}$ is well motivated, it is important to assess whether other effects could introduce biases in the DM determination. We have examined several such effects and argue that their impact on our DM measurements is negligible. First, the microshots analyzed here are broadband emissions, and thus the DM bias caused by band-limited emission is expected to be insignificant. Second, scintillation does not systematically shift the peak of the DM-S/N curve and therefore does not bias the derived DM. Third, the inferred scattering timescale of FRB 20220912A at 1.271\,GHz is approximately 0.5\,$\mu$s \citep{2023MNRAS.526.2039H}, which is about two orders of magnitude shorter than our typical sampling time. Consequently, scattering-induced temporal broadening has a negligible effect on the DM determination. We thus conclude that the DM values obtained from microshots are robust against these effects.

With a reliable estimate of DM\(_{\text{real}}\) established from microshots, we can now test whether the DM\(_{\text{model}}\) values of other narrow bursts are consistent with DM\(_{\text{real}}\). To this end, we selected nine bursts from the catalog of \cite{2023ApJ...955..142Z} that met the criteria of burst width \(< 2\)\,ms and peak flux density \(> 300\)\,mJy. These thresholds were chosen to select bursts with simple temporal profiles—where intrinsic morphological delays are expected to be minimal—and high S/N to ensure precise DM measurements.
For each of these bursts, the DM was modeled using the \texttt{DM\_phase} software package. By conducting a fine-grid search around the previously reported DM values from \cite{2023ApJ...955..142Z}, we obtained more precise DM measurements. The resulting DM values and their uncertainties, together with the original burst widths and peak flux densities, are listed in Table~\ref{tab:burst_parameters}.

\begin{table*}[htbp]
\centering
\caption{Measured parameters of the nine bursts. Columns give (1) burst number, 
(2) detection time in Modified Julian Date (MJD), (3) burst width in ms, 
(4) peak flux density in mJy, and (5) DM$_{\text{model}}$ in units of pc\,cm$^{-3}$. The MJDs, burst widths, and peak flux densities are taken from \cite{2023ApJ...955..142Z}.}
\label{tab:burst_parameters}
\begin{tabular}{ccccc}
\toprule
No. & MJD & Width & Flux & DM \\
& & (ms) & (mJy) & (pc\,cm$^{-3}$) \\
\midrule
1 & 59880.504776803 & 1.07 & $1942.5 \pm 22.3$ & $219.331 \pm 0.018$ \\
2 & 59880.505791937 & 1.71 & $1453.8 \pm 16.6$ & $219.279 \pm 0.183$ \\
3 & 59880.513993782 & 1.13 & $382.2 \pm 4.1$ & $219.379 \pm 0.040$ \\
4 & 59882.533887213 & 1.73 & $1495.4 \pm 18.6$ & $219.404 \pm 0.007$ \\
5 & 59882.540740859 & 0.63 & $2484.6 \pm 30.5$ & $219.428 \pm 0.054$ \\
6 & 59882.551230332 & 1.60 & $1088.8 \pm 13.2$ & $219.387 \pm 0.053$ \\
7 & 59882.565870588 & 1.36 & $328.3 \pm 4.0$ & $219.621 \pm 0.408$ \\
8 & 59882.567400745 & 1.57 & $1498.1 \pm 18.1$ & $219.276 \pm 0.046$ \\
9 & 59884.578511164 & 1.89 & $481.4 \pm 5.9$ & $219.587 \pm 0.181$ \\
\bottomrule
\end{tabular}
\end{table*}

\begin{figure*}
    \centering
    \includegraphics[scale=0.75]{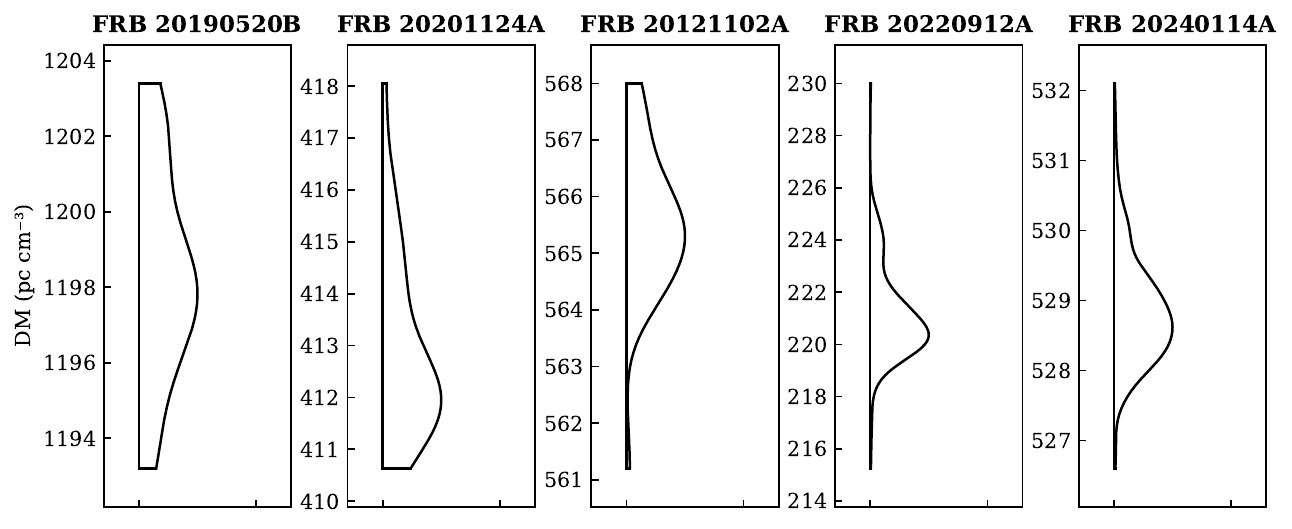}
    \caption{$\mathrm{DM_{model}}$ distributions for five repeating FRBs on their burst-active days: FRB~20190520B (MJD~59208, \cite{2026SciBu..71...76N}), FRB~20201124A (MJD~59314, \cite{2022Natur.609..685X}), FRB~20121102A (MJD~58757, \cite{2021Natur.598..267L}), FRB~20220912A (MJD~59882, \cite{2023ApJ...955..142Z}), and FRB~20240114A (MJD~60382, \cite{2025arXiv250714707Z}). Each panel shows the one-dimensional kernel density estimate of $\mathrm{DM_{model}}$ from bursts detected within a single day, including only those with DM uncertainties $<0.5$~pc\,cm$^{-3}$. The black curve traces the kernel density estimate profile over the $\mathrm{DM_{model}}$ range. In each panel, the vertical axis denotes DM (in pc\,cm$^{-3}$), while the horizontal extent encodes the normalized density amplitude.}
    \label{fig:dm_distribution}
\end{figure*}

Figure~\ref{fig:dm_vs_mjd} shows the DM\(_{\text{model}}\) values of these nine narrow bursts (blue markers), along with those of the microshots (green markers). All nine narrow bursts have DM\(_{\text{model}}\) values consistent with the weighted mean DM of the microshots (\(219.380 \pm 0.004\,\mathrm{pc\,cm^{-3}}\)) within measurement uncertainties. This agreement suggests that, for these narrow bursts, the morphological time delay \(\Delta t_{\mathrm{mor}}\) is indeed negligible (i.e., \(\Delta t_{\mathrm{mor}} \approx 0\)), making their \(\mathrm{DM}_{\mathrm{model}}\) a reliable proxy for \(\mathrm{DM}_{\mathrm{real}}\).
Given that microshots are relatively rare, the fact that narrow bursts can serve as reliable indicators of \(\mathrm{DM}_{\mathrm{real}}\) is valuable. In the absence of detected microshots, narrow bursts with simple temporal profiles offer a practical alternative for estimating \(\mathrm{DM}_{\mathrm{real}}\) in repeating FRBs.

Figure~\ref{fig:dm_vs_mjd} also shows the DM\(_{\text{model}}\) values of 766 bursts from \cite{2023ApJ...955..142Z} with DM uncertainties \(<0.5\,\mathrm{pc\,cm^{-3}}\). With the \(\mathrm{DM}_{\text{real}}\) value of \(219.380 \pm 0.004\,\mathrm{pc\,cm^{-3}}\) now established for FRB~20220912A, the observed spread in DM\(_{\text{model}}\) can be attributed entirely to \(\mathrm{DM}_{\text{pseudo}}\). The magnitude of \(\mathrm{DM}_{\text{pseudo}}\) variations reaches up to approximately \(10\,\mathrm{pc\,cm^{-3}}\).
A significant majority---705 out of 766 bursts (92.0\%)---exhibit a positive \(\mathrm{DM}_{\text{pseudo}}\), indicating a systematic bias in DM\(_{\text{model}}\) toward overestimating the \(\mathrm{DM}_{\text{real}}\). Consequently, the average DM of all bursts exceeds \(\mathrm{DM}_{\text{real}}\) by about \(1.43\,\mathrm{pc\,cm^{-3}}\), with daily mean DMs ranging from \(0.57\) to \(1.71\,\mathrm{pc\,cm^{-3}}\) above the real value.
This positive bias underscores the importance of selecting appropriate burst samples for DM analysis. We therefore recommend using the DM determined from microshots or narrow bursts---which exhibit minimal morphological delays---rather than the average value of all bursts when estimating the \(\mathrm{DM}_{\text{real}}\) of repeating FRBs.

\subsection{DM$_{pseudo}$ in other repeating FRBs}
We next examine the DM of other repeating FRBs. Figure~\ref{fig:dm_distribution} presents the distributions of \(\mathrm{DM}_{\text{model}}\) for five repeating sources on representative burst-active days: FRB~20190520B (MJD~59208; \citealt{2026SciBu..71...76N}), FRB~20201124A (MJD~59314; \citealt{2022Natur.609..685X}), FRB~20121102A (MJD~58757; \citealt{2021Natur.598..267L}), FRB~20220912A (MJD~59882; \citealt{2023ApJ...955..142Z}), and FRB~20240114A (MJD~60382; \citealt{2025arXiv250714707Z}). For each source, we construct a one-dimensional kernel density estimate of \(\mathrm{DM}_{\text{model}}\) using bursts detected within a single day, including only those with DM uncertainties smaller than \(0.5\,\mathrm{pc\,cm^{-3}}\).
All five sources exhibit a substantial spread in their \(\mathrm{DM}_{\text{model}}\) distributions. FRB~20190520B, which exhibits the most rapid DM variation among known FRBs, shows a measured decrease in $\mathrm{DM}_{\mathrm{real}}$ at a global rate of approximately $10\,\mathrm{pc\,cm^{-3}\,yr^{-1}}$ \citep{niu26}. The implied daily variation in $\mathrm{DM}_{\mathrm{real}}$ ($\sim 0.03\,\mathrm{pc\,cm^{-3}}$) is far smaller than the observed spread in $\mathrm{DM}_{\mathrm{model}}$. This suggests that true environmental changes on short timescales are unlikely to be the primary contributor to the observed spread. Instead, $\mathrm{DM}_{\mathrm{pseudo}}$ arising from intrinsic burst morphology likely accounts for most of the variation, although we cannot fully exclude contributions from other rapid propagation effects in localized environments, particularly in sources with complex local environments.

Remarkably, the magnitude of the observed spread is comparable across different repeating FRBs. For FAST observations at 1.25\,GHz, \(\mathrm{DM}_{\text{pseudo}}\) spans a range of approximately \(10\,\mathrm{pc\,cm^{-3}}\) for these sources. This consistency suggests that \(\mathrm{DM}_{\text{pseudo}}\) is a common feature among repeating FRBs and must be carefully accounted for when interpreting DM variations as signatures of environmental evolution.
We also note that \(\mathrm{DM}_{\text{pseudo}}\) may be larger at higher frequencies. For instance, at C-band, it could reach \(\sim 100\,\mathrm{pc\,cm^{-3}}\) (\citealt{reshma22}, \citealt{2025arXiv251207140W}). For FRB~20220529A, the $\mathrm{DM_{pseudo}}$ of $\sim 1-10\,\mathrm{pc\,cm^{-3}}$ in the L-band FAST data prevents the detection of gradual DM variations of $\sim 1\,\mathrm{pc\,cm^{-3}\,yr^{-1}}$ over a few-year baseline \citep{2026Sci...391..280L}. In contrast, the lower-frequency CHIME observations, where $\mathrm{DM_{pseudo}}$ is $\lesssim 0.5\,\mathrm{pc\,cm^{-3}}$, reveal a gradual DM decline of $-0.881 \pm 0.001\,\mathrm{pc\,cm^{-3}\,yr^{-1}}$ \citep{2026ApJ..1000L..53P}. This further underscores the importance of using microshots or other bursts with negligible intrinsic morphology to establish a robust baseline for $\mathrm{DM_{real}}$. 

Using our proposed method of determining $\mathrm{DM_{real}}$---i.e., employing narrow bursts as proxies---recent CHIME/FRB observations of FRB~20220912A report a marginal $2.3\sigma$ detection of a linear DM increase of $1.4 \pm 0.6\,\mathrm{pc\,cm^{-3}\,yr^{-1}}$ \citep{2026arXiv260409098A}. While this trend appears larger than the $10^{-2}\,\mathrm{pc\,cm^{-3}}$ variation assumed in A1 over multi-year timescales, it does not compromise our determination of $\mathrm{DM_{real}}$ using microshots, because Assumption A1 is only required to hold over the $\sim$one-month timescale spanned by our microshot sample, and any such long-term trend has a negligible impact on the measurements. Conversely, the CHIME long-term results further validate the effectiveness of our method in measuring $\mathrm{DM_{real}}$ and in tracking its secular evolution.
We defer a detailed determination of \(\mathrm{DM}_{\text{real}}\) for other repeating FRBs using bright narrow bursts and microshots to future work.

\section{CONCLUSIONS}\label{sec:con}

In this work, we have investigated the determination of DM\(_{\text{real}}\) for FRB~20220912A using its microshot structures. This investigation is based on two key assumptions: first, that FRB~20220912A resides in a non-magneto-ionic environment with DM\(_{\text{real}}\) variations smaller than \(10^{-2}\,\mathrm{pc\,cm^{-3}}\) over a few years; and second, that microshots possess negligible intrinsic morphological time delay, making them direct tracers of DM\(_{\text{real}}\). The main findings of this study are summarized below.

We identified two new microshots from FRB~20220912A using FAST and GBT data, bringing the total sample to four. These microshots exhibit remarkably consistent DM values over a one-month timescale, with an average of \(219.380 \pm 0.004\,\mathrm{pc\,cm^{-3}}\). This consistency confirms that microshots have negligible morphological time delay (\(\Delta t_{\mathrm{mor}} \approx 0\)), validating our assumption that the DM modeled from them directly traces the source's DM\(_{\text{real}}\).

Analysis of nine narrow bursts from FRB~20220912A—with burst widths below 2\,ms and peak flux densities exceeding 300\,mJy—shows that their \(\mathrm{DM}_{\mathrm{model}}\) values are consistent with the microshot-derived \(\mathrm{DM}_{\mathrm{real}}\) within measurement uncertainties. This indicates that narrow bursts with simple temporal profiles also exhibit \(\Delta t_{\mathrm{mor}} \approx 0\) and can serve as reliable proxies for \(\mathrm{DM}_{\mathrm{real}}\) when microshots are unavailable.

Examining the DM distributions of five repeating FRBs reveals that \(\mathrm{DM}_{\mathrm{pseudo}}\) is a common phenomenon. For FAST observations at 1.2\,GHz, the variations in \(\mathrm{DM}_{\mathrm{pseudo}}\) typically span a range of approximately \(10\,\mathrm{pc\,cm^{-3}}\) for these sources, underscoring the need to account for morphological contributions when interpreting DM variations.

These results demonstrate that microshots and narrow bursts are powerful tools for determining the DM\(_{\text{real}}\) of repeating FRBs. Accurate measurements of \(\mathrm{DM}_{\mathrm{real}}\) provide a reliable reference for interpreting burst properties and DM variations. Previous studies have shown that many repeating FRBs reside in complex and possibly evolving magneto-ionic environments. In this context, precise determinations of DM\(_{\text{real}}\) will be essential for future studies aimed at probing the physical environments and propagation effects surrounding FRB sources. We recommend using the DM modeled from microshots or bright narrow bursts as the benchmark for \(\mathrm{DM}_{\mathrm{real}}\), and advocate for long-term monitoring of FRB~20220912A to further test our underlying assumptions regarding its environmental stability and the morphological nature of its bursts.

\begin{acknowledgments}
This work is supported by National Natural Science Foundation of China grant No. 12522305, 12588202, 12203045, by the Leading Innovation and Entrepreneurship Team of Zhejiang Province of China grant No. 2023R01008, and by the Sichuan Science and Technology Program (No.2026NSFSC0743). Y.-K.Z. is supported by the Postdoctoral Fellowship Program and China Postdoctoral Science Foundation under Grant Number BX20250158. D.L. is a New Cornerstone investigator. This work made use of data from the FAST FRB Key Science Project. FAST is a Chinese national mega-science facility, operated by National Astronomical Observatories, Chinese Academy of Sciences. The Green Bank Observatory is a facility of the National Science Foundation operated under cooperative agreement by Associated Universities, Inc.

\end{acknowledgments}

%

\vspace{5mm}





\bibliography{sample631}{}
\bibliographystyle{aasjournal}



\end{document}